\documentstyle[11pt,newpasp,twoside,epsf]{article}
\def\edcomment#1{\iffalse\marginpar{\raggedright\sl#1\/}\else\relax\fi}
\reversemarginpar

\begin{document}
\title{Multi-epoch spectroscopy and {\it XMM-Newton} observations of RX J2115--58}
\author{Mark Cropper, Gavin Ramsay}
\affil{Mullard Space Science Laboratory, University College London, 
Holmbury St Mary, Dorking, Surrey RH5 6NT, United Kingdom}
\author{Tom Marsh}
\affil{Department of Physics and Astronomy, University of Southampton
Hampshire SO17 1BJ, United Kingdom}

\begin{abstract}
We present here phase-resolved optical spectroscopy and X-ray light curves of
the asynchronous polar RX J2115--58 as they change on the beat period.
\end{abstract}

\section{Introduction}

RX J2115-58 was discovered as part of the {\it ROSAT} all-sky survey and
independently by {\it EUVE}.  Further observations by Schwope et al (1997) and
Ramsay et al (1999) showed that it was an asynchronous polar with a beat period
of 6.3 days. Such a short period allows us to study the effect of a changing
orientation between the magnetic field and accretion stream on an easily
observable timescale.

\section{AAT Spectroscopy}

The left plot in Fig. 1 shows phase-folded line profiles of H$\beta$ (top), and
HeII 4686 (bottom) on the nights of 2000 July 20, 22 and 24, evenly sampling
the beat cycle (left to right). These were taken using the AAT RGO
spectrograph. The profiles are complex, with significant changes evident over
the beat cycle. An example is the enhanced heating of the secondary on the
first night, evident in the strong narrow profile in the H$\beta$ line (but not
the HeII 4686), and the enhancement at $\phi\sim0$ on July 22 and 24, which is
manifest in a brightening of the ballistic stream in the Doppler tomograms at
these beat phases.

\section{{\it XMM-Newton} Observations}

The plot on the right of Fig. 1 shows light curves in hard X-rays from our {\it
XMM-Newton} observations of this system. They consist of 7 separate
observations of two orbital periods, one each day, evenly sampling the beat
period. They are extremely variable from night to night, but these in the hard
band are not dissimilar to those seen in the {\it RXTE} X-ray light curves
(Ramsay et al 2000). We also have UV light curves from the {\it XMM Optical
Monitor}. Accretion is mainly at one pole in the lower hemisphere at beat phase
0.87, while it is clearly also accreting at a pole in the upper hemisphere at
most other phases (for example phase 0.23). The X-ray spectra also change
appreciably at different beat phases: one (at beat phase 0.13) requires a
distinct soft X-ray component while the other (at phase 0.55) does not. Similar
behaviour is seen in our shorter XMM-Newton observations of the asynchronous
polar BY Cam (Ramsay \& Cropper 2002).  A comprehensive analysis is in
progress.

\begin{figure}
\plottwo{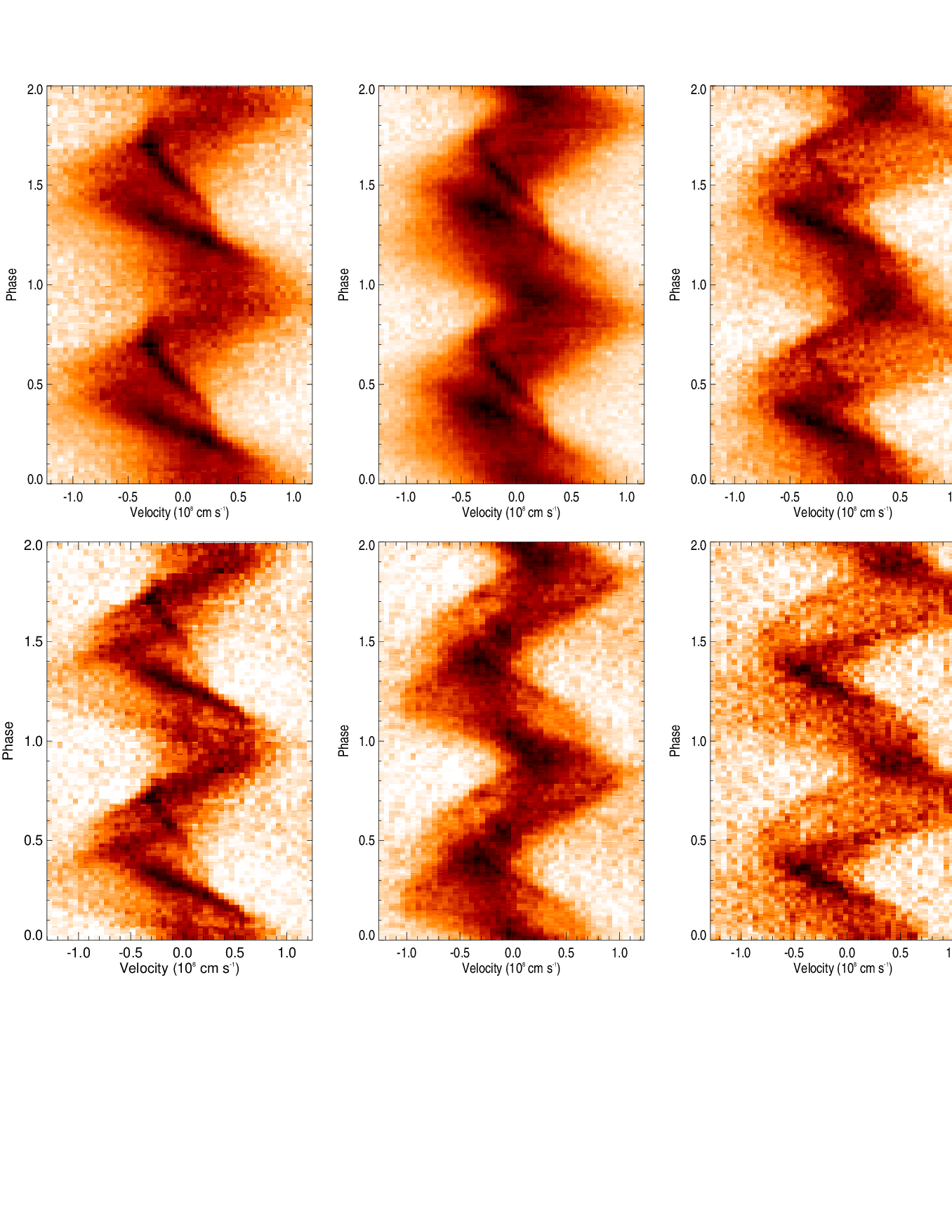}{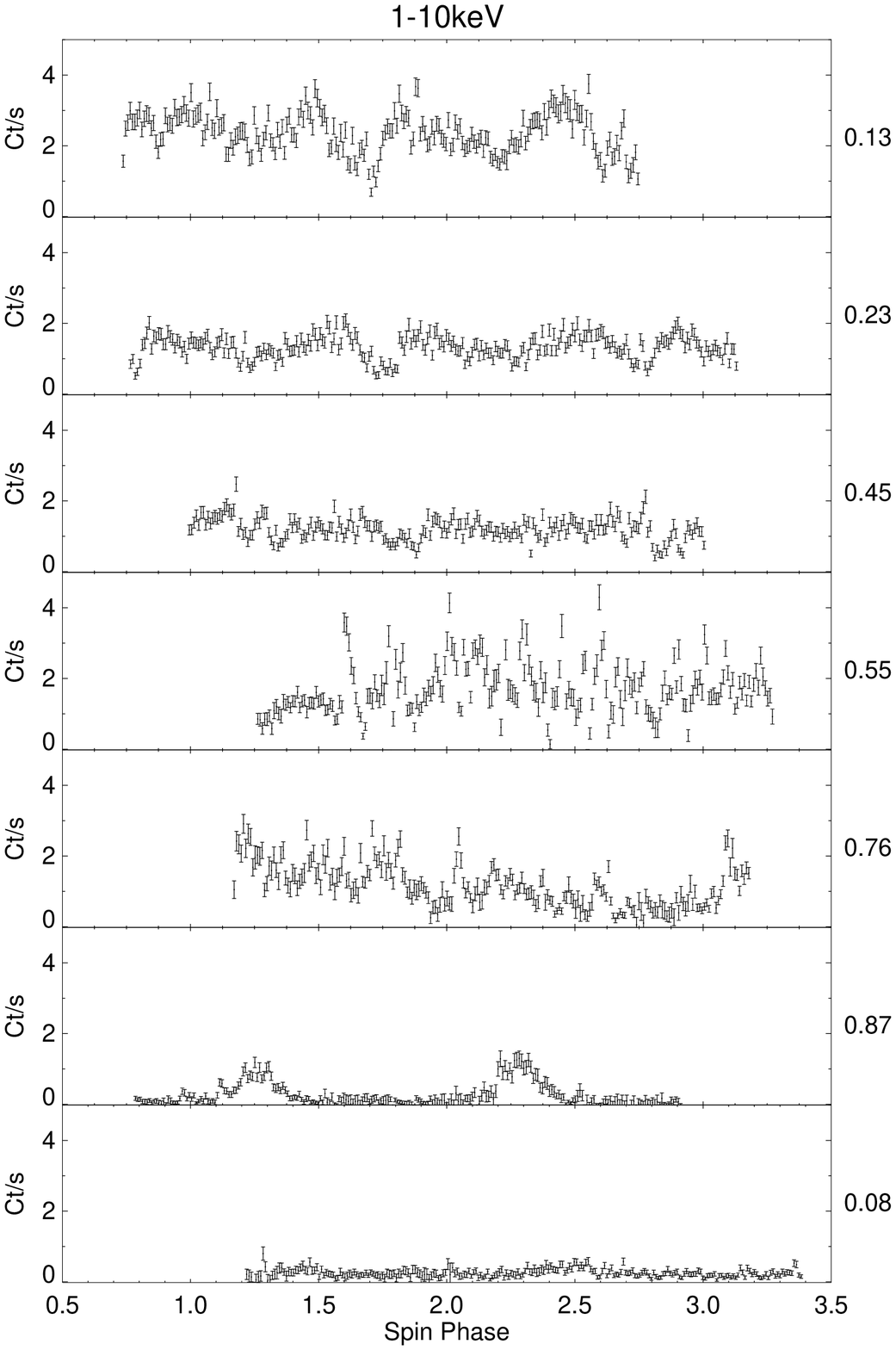}
\label{fig:folspec} 
\caption{Left: Phase folded line profiles of H$\beta$ (top) and HeII 4686 
(bottom) on each of the three nights (from left to right) July 20, 22 and 24.
Right: {\it XMM-Newton} light curves in the 1--10keV band 
folded on the orbital period for successive nights over the beat cycle. The
beat cycle phases (arbtirary zero point) are shown on the right of each plot. }
\end{figure}

\end{document}